\documentclass[sigconf]{acmart}

\AtBeginDocument{%
  }

\copyrightyear{2025}
\acmYear{2025}
\setcopyright{cc}
\setcctype{by}
\acmConference[FAccT '25]{The 2025 ACM Conference on Fairness,
Accountability, and Transparency}{June 23--26, 2025}{Athens, Greece}
\acmBooktitle{The 2025 ACM Conference on Fairness, Accountability, and
Transparency (FAccT '25), June 23--26, 2025, Athens,
Greece}\acmDOI{10.1145/3715275.3732049}
\acmISBN{979-8-4007-1482-5/2025/06}

\setcopyright{none}

\begin{document}

\title[Legacy Procurement Practices Shape How U.S. Cities Govern AI]{Legacy Procurement Practices Shape How U.S. Cities Govern AI:\\
Understanding Government Employees' Practices, Challenges, and Needs}

\author{Nari Johnson}
\email{narij@andrew.cmu.edu}
\affiliation{%
  \institution{Carnegie Mellon University}
  \city{Pittsburgh}
  \state{PA}
  \country{USA}
}

\author{Elise Silva}
\affiliation{%
  \institution{University of Pittsburgh}
  \city{Pittsburgh}
  \state{PA}
  \country{USA}
  }
\email{elise.silva@pitt.edu}

\author{Harrison Leon}
\email{hleon@andrew.cmu.edu}
\affiliation{%
  \institution{Carnegie Mellon University}
  \city{Pittsburgh}
  \state{PA}
  \country{USA}
}

\author{Motahhare Eslami}
\email{meslami@andrew.cmu.edu}
\affiliation{%
  \institution{Carnegie Mellon University}
  \city{Pittsburgh}
  \state{PA}
  \country{USA}
}
\author{Beth Schwanke}
\email{beth.schwanke@pitt.edu}
\affiliation{%
  \institution{University of Pittsburgh}
  \city{Pittsburgh}
  \state{PA}
  \country{USA}
}

\author{Ravit Dotan}
\email{ravit@techbetter.ai}
\affiliation{%
  \institution{TechBetter}
  \city{Pittsburgh}
  \state{PA}
  \country{USA}
}
\author{Hoda Heidari}
\email{hheidari@cmu.edu}
\affiliation{%
  \institution{Carnegie Mellon University}
  \city{Pittsburgh}
  \state{PA}
  \country{USA}
}

\newcommand{\eg}{\textit{e.g.,}}
\newcommand{\ie}{\textit{i.e.,}}

\renewcommand{\shortauthors}{Johnson et al.}

\begin{abstract}
Most AI tools adopted by governments are not developed internally, but instead are acquired from third-party vendors in a process called public procurement. In this paper, we conduct the first empirical study of how United States cities' procurement practices shape critical decisions surrounding public sector AI. We conduct semi-structured interviews with 19 city employees who oversee AI procurement across 7 U.S. cities. We found that cities' legacy procurement practices, which are shaped by decades-old laws and norms, establish infrastructure that determines which AI is purchased, and which actors hold decision-making power over procured AI. We characterize the emerging actions cities have taken to adapt their purchasing practices to address algorithmic harms. From employees' reflections on real-world AI procurements, we identify three key challenges that motivate but are not fully addressed by existing AI procurement reform initiatives. Based on these findings, we discuss implications and opportunities for the FAccT community to support cities in foreseeing and preventing AI harms throughout the public procurement processes.
\end{abstract}



\maketitle 

\section{Introduction}

Motivated by promises of increased efficiency, governments worldwide are increasingly adopting AI to automate bureaucratic workflows and assist  decision-making processes that impact residents \citep{reisman2018algorithmic,kim2024public,chouldechova2018case,eubanks2018automating,whitney2021hci,kawakami2024studyingpublicsectorai,levy2021algorithms}.
We situate our study in the United States, where such public sector AI applications are often not developed in house, but are instead purchased from external vendors through a process called \emph{public procurement} \cite{sloane2021ai,prier2009implications, lloyd2004public}.
In a 2023 statement, U.S. Senator Gary Peters shared that over half of the AI tools used by federal agencies were purchased from commercial vendors \cite{peters2023testimony}.
Scholars estimate that this number is even higher at lower levels of government, such as U.S. state and local governments which are less likely to have internal expertise to develop AI \citep{rubenstein2021acquiring}.
Thus, most AI systems used by U.S. governments today are developed and acquired from \emph{private vendors}.

A growing number of academic and advocacy efforts have pointed out how procured AI systems have predominantly targeted narrowly defined notions of efficiency or performance, resulting in adverse effects that disproportionately impact marginalized communities \citep{eubanks2018automating,roberts2022have,stapleton2022imagining,harcourt2006against}.
Facial recognition technologies, for example, have faced sustained criticism due to concerns about civil liberties, racial biases, and privacy violations \citep{williams2015facial,haskins2021nypd,logics2020safe}. 
Despite these concerns, in 2022, a U.S. government watchdog revealed over 20 federal contracts with private companies that either specialized in facial recognition or had awards related to its use, with a total budget exceeding \$7 million \cite{riley2022feds}.

While such incidents have exposed flaws in individual systems, they highlight deeper issues in how AI is acquired, used, and governed in the public sector.
The AI procurement process encompasses decisions of which AI tools to ask for, adopt or reject, and the manner in which they are developed and deployed: decisions of critical importance for communities susceptible to AI harms.
Scholars, civil society, and community advocates have pointed out how these procurement
decisions not only influence AI performance and risks, but also play a significant role in shaping broader governance practices and ethical standards by which AI operates in the public sector ~\cite{mulligan2019procurement,sloane2021ai,richardson2021best,rubenstein2021acquiring,hickok2024public}.

The past few years have marked an explosion of attention directed towards imagining how AI procurement might be \emph{reformed} to prepare governments to anticipate AI harms \citep{mulligan2019procurement,richardson2019confronting,conti-cookguiding,rubenstein2021acquiring,hickok2024public,govai_openletter}.
However, there is a lack of empirical research to understand how AI procurement actually \emph{occurs} in practice.
Past scholarship has established that governments must acquire AI using \emph{legacy procurement practices} that apply widely to all goods and services, including pencils, school buses, and AI ~\cite{reisman2018algorithmic,mulligan2019procurement,sloane2021ai,rubenstein2021acquiring}.
Yet, little is known about 
what exactly these practices entail, how they vary across localities, and whether governments have made any changes to their practices when assessing AI specifically. Furthermore, important information about governments' AI procurement practices is not always publicly available ~\cite{richardson2019confronting,halonen2016disclosure}.

\textbf{Our Contributions.} 
In this paper, we conduct a formative empirical study to understand U.S. cities' current procurement practices for AI.
Through semi-structured interviews with 19 local government employees in roles that involve technology governance or procurement, we pull back the curtain to show how cities' \emph{existing} purchasing practices apply to AI. 
We draw from employees' learnings from past technology procurements to identify key challenges government employees face in anticipating and addressing harms caused by procured AI systems.
We organize our findings and their implications for the FAccT community in two sections:

In Section \ref{sec:findings-one}, we offer an empirical account of how AI procurement occurs on-the-ground.
    By detailing the organizational factors that influence how procurement occurs, we show how cities' purchasing practices both enable and preclude cities' ability to address AI harms: for instance, historic norms such as cost thresholds ~\cite{duchicela2023are} allow employees to acquire low- or no-cost AI without going through the accountability measures typically associated with government purchasing. 
    We also highlight emerging steps some cities have taken to incorporate AI-specific criteria and ethical considerations into their AI procurement processes.
    
In Section \ref{sec:findings-two}, we outline three key  challenges that still remain to be addressed by cities' existing (and emerging) practices. 
We argue that more work is urgently needed to (1) address information asymmetries between governments and vendors, (2) support cities in asking more of AI vendors, and (3) support cities in sharing and assuming ongoing responsibilities of AI governance.
We discuss implications and opportunities for the FAccT community to support public sector workers in mitigating harms caused by procured AI.

Our research team has developed an accompanying white paper to this manuscript that translates our research findings into actionable recommendations for local government employees ~\cite{silva2025procuring}. The white paper is \href{https://papers.ssrn.com/sol3/papers.cfm?abstract_id=5227802}{available online (link)}.

\color{black}

\section{Related Work}

\subsection{AI in the Public Sector}
Governments' increased adoption of data-driven technologies has transformed work in the public sector \citep{alkhatib2019street,levy2021algorithms,engstrom2021artificially,stapleton2022who,madan2023ai,kim2024public}. 
While many technologies marketed as ``artificial intelligence'' today bear resemblance to historical data mining technologies of the 1990s (\eg{} to detect fraud \citep{levy2021algorithms,engstrom2021artificially}),
recent years have marked a rapid increase in the number and popularity of AI tools that employ machine learning \cite{mulligan2019procurement,levy2021algorithms,engstrom2021artificially,rubenstein2021acquiring}. 
Governments' use cases for AI, as taxonomized by \cite{engstrom2021artificially}, range from law enforcement and adjudication, such as supporting criminal sentencing \cite{forrest2021when,pruss2023ghosting}, to applications that support government service delivery, such as chatbots to facilitate communication with residents ~\cite{quinlan2024chatbot,lecher2024nyc}. 
Like private sector workers, government employees are increasingly adopting general-purpose productivity tools such as ChatGPT \citep{david2024pa,childress2024deploying,metrolab2024genai} to do the ``street-level'' work of governance \citep{alkhatib2019street}, \eg{} to adjudicate child welfare cases \cite{parmiggiani2024chapter,taylor2024ai}, thus further blurring the lines between AI's articulated scope and ultimate application.

Scholars and activists have documented the harms resulting from public sector AI systems, with the recognition that historically oppressed communities are disproportionately subject to AI-assisted punishment \citep{eubanks2018automating,stapleton2022imagining,liban2024inescapable,pruss2024prediction}. 
Public sector algorithms have faced criticism for algorithmic bias (\eg{} in the case of predictive policing \citep{bhuiyan2021lapd,capp2020primer}), lack of validity (\eg{} in the case of child welfare risk assessments \citep{coston2023validity,gerchick2023devil}), and lack of functionality (\eg{} in the case of facial identification mismatches ~\cite{hill2023eight}).
In response, a growing body of empirical scholarship has studied government employees' practices, desires, and concerns surrounding governments' use of AI systems \citep{veale2018fairness,brown2019toward,saxena2020government,stapleton2022imagining,kuo2023understanding,cheng2022how,kawakami2022improving,ziosi2024evidence}. 
Much research has focused on the perspectives of \emph{frontline workers} (\eg{} social workers \cite{kawakami2022improving}) who consume AI outputs. 
Studies have characterized how AI systems' ``misalignment" with policy objectives and workers' ethical values can result in harms to marginalized communities \citep{krafft2021action,kawakami2022improving,stapleton2022imagining,robertson2021modeling}, highlighting implications for technology design. 
Other past work illuminates the organizational factors that facilitate critical decisions about AI adoption and governance \citep{krafft2021action,veale2018fairness,kawakami2022improving,pruss2023ghosting,kawakami2024studyingpublicsectorai, kawakami2024failure}, recognizing that governments ``should not be treated as monolithic entities'' \cite{kawakami2024studyingpublicsectorai}, and that workers at varying levels of institutional power often hold varying perspectives about the appropriateness and ethical implications of AI tools.

Despite this growth in research, governments' \emph{procurement processes} remain empirically under-explored as a stage that can impact how AI is adopted and governed in the public sector.
In the following section, we describe how scholars, civil society, and activists envision public procurement as both a site of opportunity and concern for AI governance.

\subsection{Governing AI through Public Procurement}
The term ``public procurement'' generally refers to the processes governments use to bring in goods and services that are developed externally \cite{sloane2021ai,prier2009implications, lloyd2004public}, often involving paid transactions with third-party organizations.\footnote{We note that as described by past work \citep{prier2009implications}, there is no single precise agreed-upon definition for what it meant by the term ``public procurement'' -- rather, the definition is ``muddled'' and varies across contexts. See Appendix \ref{apdx:defining-procurement} for a more detailed discussion of definitions.} 
In this work, our core focus is on studying public procurement practices in the United States, particularly for \emph{municipal} (city) governments. Public administration scholars have characterized how procurement laws, organizational structures, and activities generally vary between levels of government and localities \citep{coggburn2003exploring,lloyd2004public,duchicela2023are}.

A growing coalition of policymakers, scholars, and civil society groups  have called on governments to implement AI-specific procurement guidelines \cite{uk_gov,richardson2019confronting,conti-cookguiding,rubenstein2021acquiring,hickok2024public}.
Governments must be prepared, scholars argue \citep{mulligan2019procurement,rubenstein2021acquiring,vallee2024cdt}, to interrogate the value-laden trade-offs embedded by design decisions such as the vendor's choice of training dataset, optimization function, and predictive target.
In response, several groups have explored how to embed these and other ethical considerations
within cities' existing procurement processes ~\cite{govai_openletter,richardson2019confronting,rubenstein2021acquiring,sharma2024power}.
These efforts have produced practical guidance and readily adoptable resources for government employees, such as guidelines, tools, vendor repositories, and templates to guide AI procurement practices (\eg{} \citep{richardson2019confronting,wefbox,conti-cookguiding,dotan2023how,rubenstein2021acquiring,brauneis2018algorithmic,edtech2024school}).
However, there is very little work that examines whether and how these tools help government employees in practice.

Many AI procurement reform initiatives draw from a long history of governments adapting their purchasing practices to enact social change ~\cite{rubenstein2021acquiring,grandia2017public,mccrudden2007buying}, \eg{} by creating processes that prioritize minority-owned businesses \cite{mccrudden2007buying}, enable public oversight over government surveillance \citep{crump2016surveillance,young2019municipal}, or incentivize other ethical behaviors, such as sustainability \citep{lazaroiu2020environment, varga2021how}.
These initiatives call attention to the \emph{purchasing power} governments hold ``not as regulators, but as participants'' in a free market ~\cite{mccrudden2007buying}.
Drawing from this tradition, scholars point to the ``gatekeeping role" of public procurement in deciding which AI systems are (and are not) purchased \citep{sloane2021ai,oluka2022human,conti-cookguiding,hickok2024public}. 
The ultimate hope expressed is that governments' interactions with vendors can ``change markets'' \cite{wefbox}: that governments can exercise their purchasing power to advocate for AI vendors to prioritize social values such as fairness, accountability, and transparency ~\cite{sharma2024power}.

Given this potential role of procurement in AI governance, there have been a small number of research efforts that have uncovered challenges faced by AI procurement practitioners in Kenya and Uganda ~\cite{oluka2022human} and the US federal government ~\cite{autio2023snapshot}, such as knowledge gaps in workers' understanding of AI risks or an absence of ``clearly defined standards'' for AI governance.
However, what remains less known is the current \emph{practices} that governments follow when acquiring AI, \eg{}
the assessment criteria governments apply, how  purchasing decisions are made, and who holds the power to make them.




This paper contributes to these ongoing discussions by illuminating the organizational realities in which AI procurement occurs.
We provide an empirical account of the emerging actions that U.S. local governments have taken to revise their AI procurement processes on the ground.
Given this context, in Section \ref{sec:findings-two} we revisit several critical challenges surfaced by past scholarship, 
such as knowledge gaps in government employees' preparedness to anticipate AI harms \citep{zick2024aiprocurementchecklistsrevisiting,oluka2022human,engstrom2021artificially}, governments' lack of leverage with AI vendors \citep{kawakami2024studyingpublicsectorai,hickok2024public}, and uncertainty about how to share and delegate responsibilities of AI governance \citep{aimmemo-rfi,groves2024auditing}.
We ground our analysis using examples of real-world procurements to reflect on both the progress cities have made and obstacles that remain to be overcome.


\section{Methodology}
To understand the organizational processes that shape how cities govern AI, we focus on the perspectives of municipal employees.
We conducted semi-structured interviews over six months from December 2023 to June 2024.
Our research coincided with a landmark year of government action on AI procurement, including the release and adoption of the U.S. ``AI M-Memo'' \citep{aimmemo-text,shivaram2024white}, which instituted governance requirements for federal agencies' use of AI,\footnote{The Biden AI M-Memo has since been revoked by the Trump administration ~\cite{obrien2025trump}. } and the formation of the Government AI Coalition \citep{edinger2023local,rispens2024san}, a grassroots coalition founded to ``give local governments a voice in shaping the future of AI.''
As a result, our study took place at a  moment when governments across the U.S. were just beginning to stand up and implement new governance processes for procured AI.

\emph{Recruitment.} Our recruitment process aimed to capture a diverse set of employees' perspectives on the AI procurement process, across multiple cities.
Many cities required employees to receive approval to participate in research, so we organized our recruitment efforts by city.
Once we received approval, we used snowball sampling to ask our initial contacts within a city to introduce us to other employees whose present role was involved with technology procurement or governance.
As shown in Table \ref{table:participants}, most participants worked in technology-focused roles in their city's IT or Innovation departments. 
We also spoke with specialists in vendor relations and procurement, and one HR specialist who had conducted organizational training on AI. 
Participants included both leaders who made decisions on behalf of their department, and workers whose day-to-day responsibilities involved managing AI procurement.

We began the study by reaching out to contacts in our professional networks in four U.S. cities, who introduced us to contacts in three additional cities.
From the contacts we were given, we intentionally invited cities that represented a wide range of regions and maturity surrounding AI (\eg{} whether or not they had adopted any public-facing AI policies). 
Although this approach has limitations, we learned that establishing trust through shared connections was important for government employees to feel comfortable speaking openly with academic researchers. 
For example, one participant shared that their department was hesitant to participate in research studies led by someone they were unfamiliar with. 
We invited eight total cities to participate, and seven agreed.
Participating cities represented all four major regions (Northeast, West, Midwest, and South) defined by the U.S. Census Bureau \citep{censusdiv}.

\begin{table*}[t]
\centering
\begin{tabular}{ c|c|c|c|c } 
  \textbf{ID} & \textbf{Pseudonym} & \textbf{Department} & \textbf{City Size} & \textbf{Title} \\ \hline
 P1 & Charles & Innovation & Medium & Chief Data Officer \\ 
 P2 & Mari & Information Technology & Large & Senior IT Manager \\ 
 P3 & Jennifer & Information Technology & Large & Privacy Program Manager \\ 
 P4 & Emma & Information Technology & Small & IT Business Relationship Manager \\ 
 P5 & Kai & Information Technology & Large &  Privacy Specialist\\ 
 P6 & Michael & Information Technology & Small & Director of IT \\ 
 P7 & Eric & Information Technology & Large & Chief Privacy Officer \\ 
 P8 & Liz & Management \& Budget & Medium & Sourcing Specialist \\ 
 P9 & Paul & Innovation & Medium & Innovation Specialist \\ 
 P10 & Hana & Information Technology & Large & IT Analyst \\ 
 P11 & Andrea & Information Technology & Medium & Chief Technology Officer \\ 
 P12 & Kathryn & Human Resources & Small & Talent \& Culture Program Manager \\
 P13 & Rebecca & Information Technology & Medium & Chief Data \& Analytics Officer \\ 
 P14 & Olivia & Management \& Budget & Large & Director of Procurement\\ 
 P15 & Buck & Information Technology & Large & Chief Technology Officer\\ 
 P16 & Adrian & Innovation & Large & IT Director \\ 
 P17 & Steven & Information Technology & Large & Vendor Manager \\ 
 P18 & Hugo & Innovation & Large & Chief Information Officer \\ 
 P19 & Liam & Innovation & Medium & Senior IT Manager \\ 
 \hline
\end{tabular}
\caption{An anonymous description of participating municipal employees. Titles were modified to preserve anonymity. Small cities have under 200,000 residents, medium cities have 200,000 - 500,000 residents, and large cities have over 500,000 residents. Participants were invited to choose their own pseudonym.}
\label{table:participants}
\end{table*}
\emph{Semi-structured interviews.} Following \citet{veale2018fairness}, we adopted a semi-structured interview approach to allow for flexibility in discussions.
This allowed participants to spend more time discussing the phases of the procurement process that were closest to their role and expertise.
The interviews ranged from 60 to 90 minutes and our protocol included three sections. 

First, we asked participants about their current role and work responsibilities relating to AI. 
We defined ``AI'' for participants using the OECD definition of ``any machine-based system that can make predictions, recommendations, or decisions'' \citep{oecddefinition}, and provided examples of qualifying systems. 
We chose the OECD definition to be intentionally inclusive of a wide set of predictive technologies that pose risks to residents' rights and safety, following a best practice from civil society ~\cite{vallee2024cdt}.
Second, we asked participants to walk through how an example procurement involving AI would occur in their city, paying particular attention to any differences between a standard technology procurement. 
Third, we asked participants to reflect more deeply on their perceived challenges, needs, and desires to improve the AI procurement process. 
We expressed to participants that we were interested in understanding potential risks or societal impacts of the AI systems they procured, but did not steer participants towards discussing particular types of impacts, as we aimed to understand how participants conceptualized risk and harm in their own words.
The study was approved by a university IRB.
We include our complete interview protocol in Appendix \ref{apdx:protocol} and discuss ethics and participant safety considerations in Section \ref{sec:ethics}.

\emph{Data analysis.}  We collected 23 hours of interview audio which were transcribed and coded by four team members. 
We adopted a bottom-up thematic analysis approach \citep{braun2006using} to analyze interview transcripts. 
Each transcript was first open-coded by two authors, who met to discuss each transcript and resolve any differences in interpretation \citep{mcdonald2019reliability}.
In total, we created 305 unique codes, such as ``\emph{proprietariness}'' or ``\emph{indemnification from AI harms}''. 
We then grouped these codes into themes corresponding to cities' purchasing practices, such as ``\emph{alternative procurement pathways}'', and challenges faced by employees, such as ``\emph{lack of leverage with AI vendors}''.
We complemented interview findings by conducting a document analysis ~\cite{karppinen2012what} of cities' written AI policies or procurement materials.
When available, we collected and read cities' procurement handbooks, AI policies, AI risk assessment instruments, solicitation templates, and contracting templates.\footnote{Such documents are not always publicly available, and were sometimes provided by request by our interviewees.} 
We primarily used supporting documents to corroborate findings and themes observed from interviews: for example, after learning from interviews that cost thresholds played a significant role in AI governance, we consulted each city's procurement handbook to ensure we understood the cost thresholds applicable for each city.
When selecting our final set of themes, we aimed to emphasize findings that contributed new insights to the existing literature on AI and procurement.




\emph{Limitations.} We acknowledge several methodological limitations of our study.
Our reliance on snowball sampling skewed our sample of participating cities towards large cities, whose practices and needs likely do not represent the majority of local governments in the U.S. 
While our focus on U.S. cities allowed us to draw productive comparisons across jurisdictions, we believe that understanding how procurement differs across countries and levels of government is a critical direction for future work.
Finally, our decision to focus on city employees neglects the perspectives of other important stakeholders such as AI vendors, impacted communities, and other members of the public.
We hope that our research can lay empirical foundations that can inform this future work.

\section{Legacy Purchasing Practices Shape How Cities Govern AI}\label{sec:findings-one}

\begin{figure*}[t]
\centering
\includegraphics[width=0.8\linewidth]{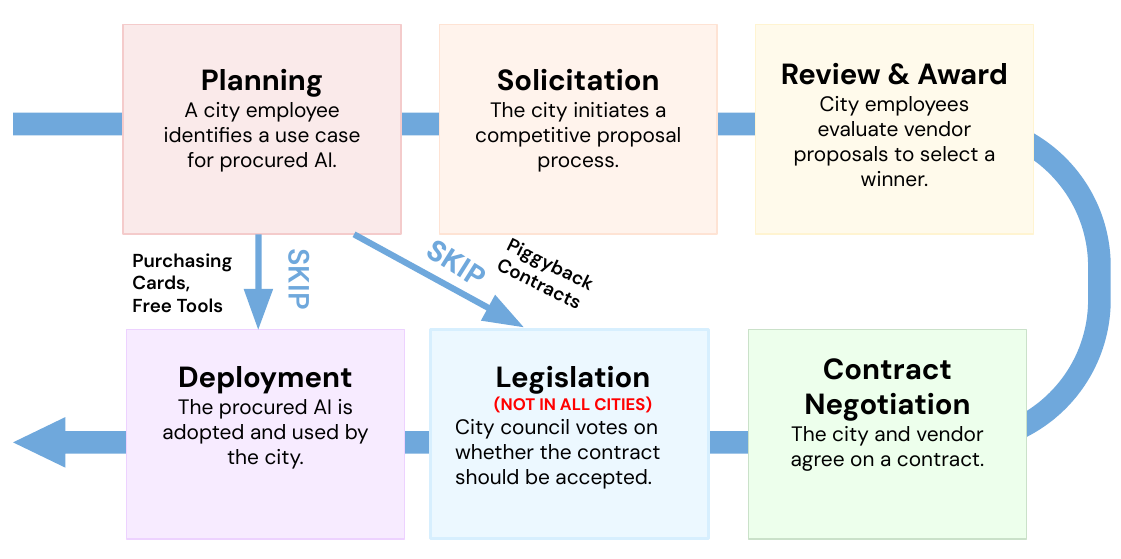}
\caption{Six common steps that occurred in cities' public procurement processes. While purchases that went through a full solicitation (\eg{} an RFP) proceeded linearly through these steps, as shown by the long blue arrow, AI acquired via alternative purchasing pathways such as cooperative purchasing agreements ("piggyback contracts") or under cost thresholds (\eg{} using purchasing cards) skipped past several steps, illustrated by the short blue arrows. In our paper, we discuss meaningful differences across cities' processes.  }
\Description[Visualization of six common steps in the public procurement process]{The figure shows an illustration of six common steps in cities' public procurement process. The steps are ordered (Planning, Solicitation, Review \& Award, Deployment, Legislation, and Contract Negotiation), which is depicted using a blue arrow pointing from the first step to the last. The figure also illustrates two alternative purchasing pathways that skip over certain steps of the process, as denoted by short blue arrows pointing to the steps that acquisitions in these pathways follow instead. }
\label{fig:six-steps}
\end{figure*}

In what follows, we shine a light on the processes that municipal employees currently use to procure AI, on-the-ground. We learned that all seven interviewed cities had already started to use procured AI technologies designed for a wide set of users and goals, for example, to facilitate resident communication, increase workplace productivity, aid law enforcement, and assist bureaucratic decision-making.\footnote{In Appendix \ref{apdx:ai-use-cases}, we describe each of these use cases in further detail and provide example real-world AI solutions mentioned by participants.} Notably, five out of seven cities we interviewed \emph{did not} have the capacity to develop their own AI solutions internally, affirming our hypothesis that local governments rely heavily (and in many cases, entirely) on third-party AI systems.

Throughout interviews, employees repeatedly emphasized that AI must go through all existing legacy procurement processes, which are shaped by a complex web of hard policy (\eg{} laws and regulations) and soft policy (\eg{} guidelines, practices, organizational culture).
We illustrate how these legacy practices shape how cities govern AI in three ways: (1) by
structuring critical decision-making, (2) determining who within a city has power to shape procurement outcomes, and (3) shaping the socio-technical values by which AI is assessed.
We conclude by summarizing the steps that some interviewed cities have taken to adapt their legacy purchasing practices for AI.



\subsection{Public procurement structures critical decisions about AI}\label{sec:practices-structure}

When asked what procurement entailed, employees often described the highly structured processes their city took to award and oversee contracts with vendors. Consistent with public administration scholarship ~\cite{curry2019contracting,halchin2006overview,financial2024essay}, employees tended to follow a six-step process (Figure \ref{fig:six-steps}) for acquisitions that involved a competitive solicitation. 
However, not all steps occurred in every city. 
While some cities required employees to introduce legislation in front of City Council to approve every procurement contract, presenting an opportunity for public oversight ~\cite{reisman2018algorithmic}, other cities did not need to introduce legislation for a purchase to be made.\footnote{See \citet{duchicela2023are} for a more detailed review of how City Council oversight processes vary across local governments.}
For each step, cities applied highly standardized processes when purchasing AI. 
Standardization, employees explained, was important to ``increase efficiency'' of the procurement process and prevent government corruption \cite{patrucco2020structuring,u4website}. 
Analyzing cities' standardized processes reveals the implicit assumptions they carry, and how existing processes both facilitate and constrain cities' capacity to address AI harms.

Throughout interviews, we learned how cities' solicitation processes determined if, and how, the societal impacts of AI factored into vendor selection. One common type of solicitation is a Request for Proposal (RFP), a structured process where a government outlines their needs, expectations, and desired outcomes. Interested vendors then submit detailed proposals that comprehensively address these requirements. After the solicitation has closed, cities review vendors' proposals to select a winner.
Liz, a procurement specialist, walked through her city's RFP evaluation process by sharing her screen as she navigated her city's procurement management software for a past IT procurement: ``So all the evaluators will show up here'', Liz said, pointing to a list of ten city employees. ``And when they go in, they score each vendor based on multiple criteria'' such as "\emph{Qualifications of the project team}", "\emph{Project plan}", and others, assigning a score of 1 to 15. Once evaluators individually had assigned scores for every proposal, they met as a committee to discuss the scores and select their ``top three candidates to reach out and set up a live demonstration''.

After a vendor is selected, the city and the vendor create a contract that specifies legally enforceable obligations for both parties. 
At minimum, contracts specify an agreed price for the city to pay over a specified time period, raising concerns of vendor "lock-in" \cite{boots2021rule} where cities are hesitant to break existing contracts. 
Steven, a procurement specialist, discussed his experiences negotiating ``indemnification'' clauses with AI vendors, which spelled out how parties would compensate each other for damages, \eg{} ``Will the vendor cover [the city] if we get sued for bias?'' Some contracts exactly spelled out the support AI vendors would provide cities: in one egregious example, a city that was struggling to reach a vendor learned that according to the contract, they were only allowed up to five phone calls of support. 

Rather than write each contract from scratch, often both the city and the vendor relied on "standard agreement" templates written by lawyers. 
Vendor's standard agreement templates written for government customers included detailed \emph{terms-of-service} \cite{fiesler2016reality} for the technology and the support services the company would provide. 
All interviewed cities had created their own standard agreement templates for IT procurements to ``protect the interests of the city'' by laying out clear expectations of vendors, \eg{} that the vendor must comply with minimum data security standards, or provide the city ownership over any data collected from the engagement.

Uncovering cities' standardized processes reveals how critical decisions about AI are currently structured, and how efforts to mitigate algorithmic harms might be operationalized within existing processes, \eg{} via solicitation questions, scorecards, and contract terms. 
However, a critical examination of these practices surfaces how they might limit cities' ability to effectively address AI harms. For example, FAccT scholars have documented the limitations of relying on highly structured methods (\eg{} compliance "checklists" ~\cite{madaio2020codesigning,wong2023seeing} that are highly similar to RFP requirements) to assess the societal implications of AI. 
Similarly, FAccT research has highlighted the difficulty of anticipating algorithmic harms before deployment \cite{boyarskaya2020overcoming,shen2021everyday,shelby2023sociotechnical,moss2021assembling}, arguing that the socially situated nature of harms arising from lived experiences mandates their assessment "from the ground up" ~\cite{moss2021assembling}. 
However, cities were required to negotiate and sign contracts with AI vendors before they could access (and deploy) AI systems behind vendor paywalls. 
Contracts were often not renegotiated, and indeed, several interviewed employees [P2, P4, P17] regretted how they negotiated a particular contract only after algorithmic harms surfaced post-deployment. 
Thus, procurement contracts codified both cities' continued financial obligations to vendors, and vendors' (lack of) obligations to redress algorithmic harms.

\subsection{Public procurement determines who has visibility, oversight, and decision-making power over AI}\label{sec:practices-people}

Cities' legacy procurement practices often specified accompanying roles, responsibilities, and lines of communication for employees. Many cities conceived of procurement as a highly collaborative activity involving not only the purchasing customer, but also specialists from a wide variety of departments: procurement specialists to assist with solicitations, legal specialists to help with contracting, and IT specialists to provide expert review. 
Different employees who participated in an AI procurement had different vantage points and objectives depending on their roles. 
However, not all of these specialists were involved, or even aware of, every acquisition. Differences in cities' established purchasing processes shaped \emph{who} within the city had visibility, oversight, and decision-making power over procured AI.

While all cities had designated IT staff whose responsibilities included providing support on technology procurements, cities' procurement processes differed in how they allocated decision-making power. 
Some cities established IT reviews as mandatory approval gateways, but other cities positioned IT reviews as optional consulting. For example, because Adrian's city historically granted individual departments (\eg{} the police, public schools) ``a lot of autonomy in how they operate'', city employees in these departments were not required to consult IT before making their own technology purchases. 
In contrast, Jennifer's city mandated that all city employees, regardless of department, must ``open a ticket'' to initiate an IT review before buying new software.

Another key factor that determined the actors involved in an AI procurement was the purchasing pathway used to make the acquisition. 
In Figure \ref{fig:six-steps}, we describe the six steps that occur for procurements that go through a full solicitation.
However, our interviews indicate that AI procurement often does not take this classic route, sometimes skipping competitive solicitation or contract negotiation entirely. Instead, employees leveraged a variety of \emph{alternative procurement pathways} to purchase AI. 
In many cities, acquisitions that occurred through these alternative procurement pathways ``didn't have to go through a procurement [department]'', and thus fell outside the scope of existing accountability measures for government purchasing.
Due to broader shifts in the AI landscape, namely the availability of low- and no-cost AI tools, many AI acquisitions did not involve a competitive solicitation because they were under specified \emph{cost thresholds} \cite{duchicela2023are} that would require going to solicitation. 
Local policies specified that municipal employees could make purchases under a certain dollar amount (which varied city-to-city) at their own discretion, using a government-issued \emph{purchasing card}.
Cost threshold policies, procurement director Olivia explained, were established to ``triage risk, by defining the [riskiest projects] based on dollar value''. 
Types of AI tools that fell under cost thresholds included free online services like ChatGPT, services with paid subscription models, or AI donated through academic collaborations, foundations, or from for-profit companies.\footnote{For example, Palantir, a for-profit company, donated its predictive policing technology as a ``philanthropic gift'' to be used by the City of New Orleans, free of charge ~\cite{stanley2018new}.}

The governance processes applicable to procurements under cost thresholds varied significantly from city to city. Adrian, an IT leader, reflected on how technology acquisitions under cost thresholds were particularly difficult for their department to even be aware of until after they were already purchased: ``[For purchases] below \$x0,000, there's few oversight of regulatory mechanisms to control, or even have visibility of what departments do,'' Adrian explained. ``It's not routed through a centralized control mechanism.'' As a result, individual employees could purchase and deploy free and low-cost AI tools (\eg{} with prices under \$x00 per month) at their own discretion, without notifying others in the city.
In contrast, Mari's city still routed "anything involving a company or vendor" through a centralized procurement review: ``Even if it's an in kind donation, it is still tied to a scope-of-work and a contract.'' 

Cities' varying governance processes for different types of AI procurements had significant impacts on individual employees' ability to advocate for their desired outcomes: for instance, refusing to proceed with procuring an AI system due to concerns about its potential harms.
In some cities, employees who were tasked with doing AI governance work often lacked basic visibility into the AI their colleagues were using due to these legacy purchasing norms. 

\subsection{Public procurement shapes the socio-technical values of AI assessment}\label{sec:practices-values}
In conversations, we found significant differences in the socio-technical values that shaped cities' approaches to assessing AI systems. 
These values, we found, were rooted in cities' existing public procurement norms and established technology review processes.

Public procurement norms, particularly cost-effectiveness, played an important role in how employees assessed AI proposals.
``When we make decisions, we think about how we're stewards of residents' tax money,'' explained Emma, an IT procurement specialist.
Indeed, cost was a central factor that influenced employees' AI purchasing decisions.
Employees shared many anecdotes where they ultimately decide not to procure an AI system because it was too expensive [P4, P6, P8, P14].

On the other hand, some AI technologies' low price tag, or promised cost-saving capabilities, served as an incentive to some.
While employees like Hugo expressed enthusiasm about the increased availability of low-cost consumer AI tools (such as a generative AI pilot that only cost Hugo's city \$5), others expressed concern over the ``extractive capabilities'' and hidden costs of free AI tools.
Adrian recalled an incident where an employee used a transcription AI tool that did not have ``a consensual model for data collection''.
Although the transcription tool was free, its use could result in ``divulging resident information'' or ``more secure'' government business, data that the vendor could use to improve its models.
Adrian's story highlights a tension between the long-standing norm of cost-effectiveness and the new reality of unique risks posed by low- and no-cost AI.

Beyond cost, employees also applied criteria from their established IT review processes when purchasing AI. 
Understanding these IT criteria is important because in many cities, IT review processes were the only applicable governance processes for procured AI. 
IT personnel were responsible for assessing review criteria that varied across cities, often operationalized as "technical" or "functional" requirements in RFPs. 
Common review criteria included the software's security (\eg{} to data breach attacks), usability, ease of set-up, and compatibility with the city's existing technology stack. 
In conversations, we found that these common criteria for software procurements were insufficient to anticipate AI harms. 
For instance, basic considerations of system functionality ~\cite{raji2022fallacy}, validity ~\cite{coston2023validity}, or train-test distribution shift ~\cite{quionero2009dataset} were not typically asked during a traditional software procurement, where systems are often assumed to be ``programmed correctly'' to achieve specified goals. 



One significant difference between cities' IT review processes was the degree to which they prioritized values such as data privacy and protection from government surveillance.
Although these considerations were never mentioned in most employees' descriptions of their review criteria, participants in a small number of cities called them out as organizational priorities. 
Understanding these priorities, we learned, was critical to understanding how cities procured AI. 
For example, cities with established privacy programs often tasked privacy personnel to work on AI-focused initiatives, and adapted existing privacy governance infrastructure to include AI-specific considerations.
Similarly, cities with surveillance ordinances had already established governance processes such as community oversight mechanisms for qualifying AI technologies ~\cite{young2019municipal}.

Organizational values also shaped the way employees conceptualized AI's benefits and risks: Kai shared that it was the "privacy lens" adopted by her organization that led her to ask vendors questions about training dataset curation and consent. 
In contrast, employees in cities without surveillance or privacy policies spoke considerably more positively about AI-powered ``smart city'' ~\cite{whitney2021hci} or policing technologies.
Thus, cities' (lack of) existing privacy or surveillance policies played a critical role in determining what risks were considered, what mitigations were requested, and what technologies were ultimately purchased and deployed.

\subsection{Emerging practices: New actions cities have taken to govern AI}\label{sec:new-practices}

Three interviewed cities had already implemented specific changes to their existing procurement practices for AI, and several other city leaders shared their future plans to change their practices. City employees felt that it was "early days" in revising their AI governance processes: for example, employees were in the midst of overseeing their first AI solicitation, conducting their first AI risk assessments, and revising their AI governance processes more broadly. 
With this rapidly evolving landscape in mind, we highlight important changes that cities have made so far.

\emph{Intervening within existing procurement processes.} The most common action considered by cities was to modify their existing contract and solicitation templates to include additional items for AI-related procurements.
Cities instituted additional reporting requirements for AI systems in their technology RFPs, adopting language from  the GovAI Coalition's AI FactSheet ~\cite{govai_factsheet} of questions for vendors.
The factsheet asks vendors to report ``essential technical details'' such as on what data the AI was trained, under what conditions the system was tested, the values of relevant performance metrics, and steps that vendors have taken to promote values of fairness, robustness, and explainability.
Cities also adapted their contracting practices by developing additional contract language specific to AI procurements. 
Many employees adopted language from GovAI's Vendor Agreement ~\cite{govai_vendoragreement}, a ``plug-and-play legal addendum'' of requirements for vendors, \eg{} to develop an AI incident response plan, remediate AI incidents, and provide governments with a means to monitor and audit AI performance. 
Most interviewed cities did not plan to adopt GovAI's resource templates as-is, but instead chose to ``selectively adopt'' the items that felt most important to their city.



Cities' legacy purchasing practices shaped how applicable these procurement interventions were to different AI systems. For instance, AI acquired under cost thresholds did not go through a competitive solicitation or contract negotiation, leading the cities to accept vendors' terms-of-use as is. 
Similarly, AI acquired by ``piggybacking'' by adopting an existing contract between the vendor and another government ~\cite{kilka2022abcs} often did not involve re-negotiation to include additional terms. 
As a result, these interventions did not apply to the many AI solutions acquired through these other pathways. 
In recognition of the limitations of relying exclusively on existing procurement infrastructure for solicitations and contracting, several employees decided to organize their AI governance efforts around a separate "AI review".

\emph{Establishing an AI review process.} 
Three interviewed cities established an internal AI review process overseen by IT employees, that occurred outside of the formal review processes of procurement (\eg{} scorecards).
When a city employee wanted to acquire an AI solution, they initiated an AI review by contacting these IT specialists directly, \eg{} by filling out a form or ``opening a ticket''. 
IT employees or committees trained to assess AI systems then reviewed the employees' request.
AI reviews often involved conducting \emph{risk assessments} to help city employees understand potential impacts of the system.
Employees used pre-acquisition risk assessments to shape subsequent conversations with AI vendors, for instance, by requesting specific risk mitigation steps to be implemented in the AI system's design.
Some cities used risk assessments to institute additional governance steps for ``high-risk'' systems, such as instituting internal usage protocols, public reporting requirements, and regular post-deployment monitoring.
We discuss governments' emerging AI risk assessment practices in detail in Appendix \ref{apdx:risk-assessments}. 

Separating AI reviews from cities' formal procurement processes had several benefits, such as improving process efficiency so that employees could contact expert AI reviewers directly.
Cities could require AI reviews for acquisitions that would not typically be reviewed by their city's procurement departments, such as free or low-cost AI tools.
However, employees' efforts to operationalize and socialize new AI reviews in practice often relied heavily on existing procurement infrastructure.
For instance, Hana's city implemented their AI review within their city's online procurement system that had existed for years.
This centralized procurement system, Hana explained, ``allows us to be pretty confident that if there is some kind of AI system, we're going to catch it, because it's going through this procurement process''.
But this level of centralization was uncommon across other interviewed cities, where IT employees only became involved once a formal solicitation was initiated.
Thus, the feasibility and implementation of cities' AI reviews was still heavily influenced by their legacy procurement practices.


\vspace{-7pt}
\section{Three Key Challenges at the Heart of AI Procurement Reform}\label{sec:findings-two}
Although some employees pointed us to success stories where their purchasing processes enabled them to walk away from red flags or negotiate for mitigations, we found that cities' present AI procurement practices often fell short of their ambition to protect the public from risks and harms of AI.
In what follows, we reflect on three challenges that often motivate, but are not adequately addressed by existing AI procurement reform efforts.
We revisit challenges previously surfaced by past scholarship ~\cite{veale2018fairness,engstrom2021artificially,autio2023snapshot,kawakami2024situate,hickok2024public} to reflect on the considerable progress that has been made, and the work that is still urgently needed to (1) address information asymmetries between governments and vendors, (2) support cities in asking more of AI vendors, and (3) support cities in sharing ongoing responsibilities of AI governance.
Under each challenge, we identify opportunities for how the FAccT community might contribute towards addressing local governments' needs for support.



\subsection{Information asymmetry and (un)preparedness for AI}
While some employees expressed confidence in their organization's preparedness to govern procured AI, many employees, particularly in smaller cities, felt behind. Employees of cities that had not yet introduced AI standards into their procurement processes were often uncertain of ``what questions to ask'' AI vendors. One procurement specialist reflected on how her unfamiliarity with AI made her feel: ``I don't really know what the risks are to working with AI. If I can't protect us from those risks comfortably, then I'm not doing my job.'' 

This information asymmetry, defined by ~\cite{bloomenthal2024asymmetric} as a phenomena ``where one business party possesses more information than the other party they are dealing with'', resulted in procurements where important questions of system functionality, training data provenance, or other societal impacts were never raised. 
Sales representatives, who held significant sway in educating governments about AI use cases through pitch-like demos, shaped cities' narratives about procured AI technologies' potential benefits and harms. We learned from several anecdotes that AI vendors did \emph{not} report basic information about AI systems unless asked. 
For instance, one participant shared that their team did not consider risks posed by hallucinations (providing incorrect information) in a recent procurement of an AI chatbot service, and that risks due to inaccuracy were ``not part of the conversation'' they had with the vendor.

Employees shared a variety of methods they used to increase their organization's AI literacy and preparedness, such as recruiting employees with AI expertise, resource sharing efforts (\eg{} through GovAI), and participating in intergovernmental AI task forces ~\cite{dwyer2025state} or grassroots peer networks. 
Some employees began collaborations with academic researchers, for example, to evaluate a pilot of a generative AI productivity tool.
Finally, some cities hired paid consultants (\eg{} from Gartner) to provide strategic advising (\eg{} conduct market research) on AI. 
Employees in smaller cities were less likely to be aware of existing resources or peer networks to learn about AI, and felt overwhelmed by the ecosystem of ``hype cycles'' surrounding emerging generative AI technologies.

While existing resources and standardization efforts helped cities identify questions to ask of vendors, employees repeatedly struggled to understand what to do with the information that vendors provided in response. 
Jennifer, a privacy specialist, shared how her department was uncertain about how to interpret the performance metrics reported by vendors: ``We ask some sort of question: what is your R-squared value? How do we know if [what is reported] is good? Someone needs to be able to have the technical acumen to say what is acceptable here in terms of accuracy, error rates, thresholds.''
While employees like Jennifer, Eric, and Paul expressed a similar desire for clear guidance and thresholds, existing AI procurement reform initiatives intentionally left these decisions to the discretion of individual cities, who could determine their own organizational risk tolerance. 
However, even cities that knew what to ask still struggled to individually assess and make decisions based on information reported by vendors.

\paragraph{Recommendations for future work} Although much progress has been made in developing resources to support governments' AI procurement practices, our findings reveal that there is still an urgent need for government education and support.
Misconceptions surrounding AI capabilities, often perpetuated by AI vendors, resulted in power imbalances where cities' existing review criteria for software proved insufficient to anticipate AI harms.
We believe that the FAccT community has a critical role to play in doing this translation work.
Our conversations revealed a number of avenues through which members of the public can influence AI policymaking, such as by collaborating on research ~\cite{stapleton2022who}, joining peer networks such as the GovAI Coalition ~\cite{govai_openletter}, or even simply reaching out to one's own government to understand their AI governance practices and needs for support.
Following interviews, our research team curated and shared educational resources that we hoped would address employees' open questions about AI.
These conversations, we learned, were often a precipitating event that led interviewed cities to join AI-focused peer networks, revise their procurement practices, and even establish new ethical AI governance policies.
The FAccT conference can explore models to build community and develop trust-based relationships across scholars, community advocates, and government employees,
\eg{} following models such as the Public Tech Leadership Collaborative ~\cite{ds_ptlc}.

Future research should also explore tools and processes to help governments assess information provided about AI systems, to make decisions about \emph{what to purchase}.
Our conversations surfaced several obstacles to measuring pre-acquisition performance. 
A lack of standardized benchmark datasets for public sector use cases meant that vendors reported metrics using their own proprietary datasets, making comparison across vendors difficult.
More broadly, participants expressed a desire for vendors' AI systems to be vetted \emph{for them}, for instance, through a third-party certification program ~\cite{cihon2021ai,costanza2022who}.
Future efforts can aim to provide domain-specific guidance on what measures should be reported, develop standardized benchmarks, and explore the possibility of third-party certification, with a critical recognition of where the narrow framing of performance measurement might fail to capture the ethical implications of AI adoption ~\cite{moss2021assembling,geiger2023rethinking,johnson2024fall}.
We join past calls ~\cite{young2019municipal,moss2021assembling,kawakami2024studyingpublicsectorai,hickok2024public} for more empirical research that examines how members of the public (\eg{} impacted communities and their advocates) can have a ``meaningful opportunity to respond and, if necessary, dispute the use of a particular AI system'' before acquisition occurs ~\cite{reisman2018algorithmic}.

\subsection{Challenge: Cities lack leverage in relationships with AI vendors.}
Several employees from cities with new AI governance measures described positive experiences with "good vendors" who were transparent and responsive. For instance, an AI vendor's openness to disclose system limitations in their autonomous drone technology helped a city refine its usage policies to avoid its deployment in low-accuracy environments.
Despite these successes, we heard many more stories of uncooperative AI vendors that failed to provide basic information about their AI system, amend their contracting terms, or implement mitigations requested by the city. 
We found that employees repeatedly felt that they lacked leverage in advocating on behalf of their city throughout the procurement process.

AI vendors frequently withheld critical information about their AI systems from governments, hindering employees' ability to assess risks. Vendors often invoked intellectual property claims to deny basic requests, such as responding to AI RFP items or completing AI FactSheets. 
Consistent with past research ~\cite{brauneis2018algorithmic,sloane2021ai,mulligan2019procurement,kawakami2024studyingpublicsectorai}, participants were often denied access to vendors' AI models (\eg{} model weights) and training datasets. Additional denied requests included information about the presence of copyrighted content in training data [P3, P5], whether city data would be used to train vendor's models [P1, P17], and asks for disaggregated accuracy measures, \eg{} across demographic groups [P2, P7, P10].

While bad vendor behavior could serve as an informative signal when making purchasing decisions, relational challenges still arose for procurements that were already under-contract. Past procurements were important to consider given the prevalence of vendors' ``scope creep'' ~\cite{stevens2022how} that introduced AI under existing contracts. 
Employees shared anecdotes where AI vendors did not provide adequate personnel training [P4, P17], refused to give cities access to data collected on city employees or residents [P2, P14, P15], and refused to let cities opt out of new AI features [P1, P17] or their data being used for AI training [P5, P13, P16]. In some cases, cities responded by ``shutting off'' purchased services, such as a city that stopped paying for Zoom because they were ``uncomfortable with their AI notetaker.'' 
Yet, several employees discussed past procurements where they struggled to break ties with vendors.
For instance, employees shared stories where they struggled to disable AI features that were deeply embedded in necessary technical infrastructure, such as new AI productivity features introduced into enterprise software [P1, P3, P5, P6, P11, P15]. 

More broadly, we observed that employees' optimism about their city's ability to hold vendors to account was shaped by the decision-making leverage they held within their own city. When vendors failed to meet employees' expectations or presented them with take-it-or-leave-it scenarios, employees who felt they had the authority to influence purchasing decisions felt comfortable walking away. For instance, Rebecca, an IT director, worked in a city that had passed a policy requiring AI reviews as a necessary approval workflow. When an AI vendor told Rebecca that basic information about their model's performance was ``proprietary'' (an ``unacceptable answer'' according to city policy), Rebecca made the decision on behalf of the city to not make the purchase. 
But in other cities, employees who oversaw cities' AI review processes lacked the decision-making power to influence final purchasing decisions. As a result, many AI procurements where vendors failed to meet cities' standards still ultimately moved forward.

\paragraph{Recommendations for future work} 
We found that many AI vendors refused to voluntarily comply with cities' requests to provide basic information or implement simple harm mitigation steps. 
This context of low compliance is critical to consider in this moment of unprecedented attention targeting governments' AI procurement practices. 
Several interviewed employees expressed their frustration in having to advocate for basic ethical behaviors (\eg{} transparency about how the vendors would use resident data) that they felt should be minimum requirements to even be able to participate in the market (\eg{} enforced as consumer protection rights or through federal data privacy legislation).
These stories call into question the hope implicit to initiatives that focus primarily on governments' power \emph{as purchasers}: that when it comes to protecting the public interest, the market of public sector AI vendors will simply regulate itself.



When vendors failed to meet expectations, a variety of factors such as contract agreements or conflicting incentives across employees meant that procurements still moved forward.
Future research efforts can more deeply interrogate the role that legacy purchasing practices play in constraining cities' negotiation power, and ways that cities might adapt their processes to address these issues head-on, \eg{} by explicitly giving AI reviewers the power to reject a proposed AI system due to its potential to cause harm.


\subsection{Sharing (and assuming) ongoing responsibilities of AI governance}
Procurement agreements create a unique type of accountability relationship: governments established to protect public values must rely on vendors to develop AI solutions. But the work of AI governance does not end at development, or even deployment. Review and oversight of procured AI technologies requires labor and expertise to conduct evaluations, respond to incidents, train users, and enforce compliance. 
In this section, we share cities' experiences working with AI vendors to evaluate system performance to explore cities' varying perspectives on delegating AI governance work.

In our conversations, we found that vendors often did not provide performance evaluations after deployment. In response, some participants [P7, P10, P15, P18] evaluated models ``on their own''. For example, Hana's department sought to monitor the effectiveness of a procured gunshot detection AI system to ensure it ``was continuing to be effective and meet the needs of the city.'' However, the vendor neither shared performance evaluations nor offered guidance on how to assess the system. As a result, Hana and her colleagues had to ``develop their own metrics'' and monitor performance themselves. 

Some employees embraced the opportunity to conduct independent evaluations. Hugo, for example, valued the city's ability to shape evaluation criteria both as a safeguard against vendor's selective reporting and as an opportunity to strengthen local AI expertise. However, not all employees conducted their own evaluations of procured AI systems, and many participants [P2, P3, P4, P5, P8, P13, P17] believed that conducting performance monitoring and audits should be the responsibility of the vendor. 

Considering performance evaluation as a case study reveals several insights for understanding how cities and vendors can share responsibilities of AI governance at large. Employees pointed out how conflicting incentives (\eg{} companies' unwillingness to report unfavorable results) might actually dis-incentivize vendors from fully participating in activities of governance – in this case, evaluating system efficacy. While employees who took ownership of doing evaluation work often found it to be an rewarding opportunity to promote the best interests of their city, other employees expressed hesitancy in their ability and capacity to do AI governance work.

\emph{Recommendations for future work.} 
To date, there is a lack of guidance on how the ongoing responsibilities of AI governance should be shared between cities versus vendors.
Although several city employees expressed concerns about their capacity, we believe that
most AI governance activities are \emph{best performed} by government employees, who are better positioned to represent the needs of city employees and residents throughout this process (\eg{} in defining what is most important to measure ~\cite{jacobs2022hidden}).
We join past work in arguing that internal capacity-building is also critical for establishing robust systems of \emph{internal} accountability and ownership over procured AI ~\cite{reisman2018algorithmic,engstrom2021artificially}.
Thus, future research can expand on existing tools and processes to support governments in various aspects of AI governance, \eg{} training users of AI systems ~\cite{kawakami2023training} and evaluating system performance ~\cite{kuo2023understanding}.
To be effective, interventions should be developed with both the unique context and constraints of procurement in mind, such as employees' non-technical expertise and capacity, as well as constraints employees face in accessing vendors' ``proprietary'' AI models ~\cite{casper2024black}.

\section{Conclusion}
In this paper, we shed light on the emerging AI procurement practices of seven U.S. local governments.
We illustrate how legacy purchasing norms that apply to \emph{all} goods and services -- school buses, desktop computers, and AI technologies -- shape how AI is assessed, selected, and governed on the ground.
We summarize key takeaways from our research that contribute to the growing literature on AI procurement:
\begin{enumerate}
    \item \emph{Local governments' procurement practices vary substantially and meaningfully across jurisdictions}. Cities' \emph{legacy} procurement practices that are \emph{applicable to AI} vary along the three dimensions we identified in Sections \ref{sec:practices-structure}, \ref{sec:practices-people}, and \ref{sec:practices-values}, with significant implications for AI governance. 
    This matters because past work on AI procurement often paints a simplified, unified view of how local governments' procurement processes unfold (\eg{} ~\cite{richardson2021best,rubenstein2021acquiring,wefbox}).
    While taking a broad view allows one to make widely applicable recommendations, when it comes to AI procurement, the devil is in the details -- of cost thresholds, purchasing pathways, and who is (and isn't) brought in when making purchasing decisions.
    We call on researchers to undertake further \emph{comparative studies} that examine how AI procurement norms vary across local and state contexts. 
    By drawing on lessons from their peers, governments can identify actionable pathways to adapt or reform their own procurement practices.
    \item \emph{AI technologies are most often} not \emph{acquired via a formal acquisition process} (\eg{} using a competitive solicitation/RFP). 
Instead, most AI technologies are acquired via alternative \emph{purchasing pathways} that often evade traditional oversight mechanisms for government purchasing, such as purchasing cards, free subscriptions, donations, or piggyback contracts.
    Most efforts to reform governments' AI procurement practices target the formal solicitation process ~\cite{wefbox,brauneis2018algorithmic,conti-cookguiding,dotan2023how}, and thus are not applicable to most real-world AI acquisitions.
    
    \item While governments are still in early stages of adapting their AI procurement practices, our conversations reveal a deeply concerning trend: \emph{many AI vendors are not cooperating with employees' efforts to understand and mitigate AI harms}.
    The implications of vendors' non-compliance for the rights and freedoms of vulnerable communities cannot be understated as local, state, and federal governments increasingly outsource the critical work of governance to third-party AI systems ~\cite{mulligan2019procurement,liban2024inescapable,hickok2024public,booth2025mainlined}.
    Efforts to support cities in navigating these power differentials, \eg{} by conducting their own evaluations, are urgently needed. 
\end{enumerate}


Looking forward, we encourage the FAccT community to reflect on what intervening within the structures of public procurement affords, and its limitations as a remedy for algorithmic accountability.
Our study prompts us to ask: what types of reforms to cities' legacy purchasing processes might the FAccT community advocate for?
What happens when cities' negotiations with AI vendors fail to be successful? 
When might we look towards other approaches, such as direct regulation of AI vendors ~\cite{warburton2024us,kang2024states}, or proposals that shift meaningful decision-making power towards members of the public ~\cite{reisman2018algorithmic,moss2021assembling}, 
to protect the public interest?

\newpage
\section{Endmatter Statements}


\subsection{Ethical Considerations}\label{sec:ethics}
To preserve anonymity of participating employees and cities, we assured interviewees that their participation was voluntary, they could decline to answer interviewer questions, and their responses would be kept anonymous. 
For sensitive or potentially identifying interview quotes, we exclude participant IDs to preserve anonymity. 
When appropriate, we use the ``x'' character to omit exact dollar amounts to preserve confidentiality.
To mitigate the risk that participating cities are identified, we limit the amount of detail we provide about each cities' practices.
Participants' department and job titles were modified to reduce the risk of re-identification.
\begin{acks}

We thank the municipal employees who participated and made this research possible.
We thank members of the San José Government AI Coalition and MetroLab Network for formative conversations that shaped this work.
We thank Samantha Dalal, Anna Kawakami, members of the Heidari Lab, attendees of Carnegie Mellon's Fairness, Ethics, Accountability, and Transparency reading group, and our anonymous reviewers for offering feedback on our manuscript.

This project was supported by the Responsible AI Seed Fund at Carnegie Mellon's Block Center for Technology and Society. 
NJ and HH acknowledge support from the NSF (IIS2040929 and IIS2229881) and PwC (through the Digital Transformation and Innovation Center at CMU). 
Any opinions, findings, conclusions, or recommendations expressed in this material are those of the authors and do not reflect the views of the National Science Foundation and other funding agencies.
\end{acks}

\bibliographystyle{ACM-Reference-Format}
\bibliography{sample-base}

\newpage
\appendix
\section{Defining ``procurement''}\label{apdx:defining-procurement}

All the definitions of "public procurement" that we reviewed encountered share fundamental similarities: they all concern the process of bringing in goods and services that are developed externally, to achieve the goals of a public sector entity. 
They differ, however, in specific components of the process. For example, while the United States’ federal definition emphasizes a “competitive” purchasing process, denoting the exchange of money as part of procurement, some local governments, like New York City, have definitions that are broader, encompassing all functions related to obtaining goods and services whether or not money changes hands \citep{aquisition2024part,NYC2024glossary}.

In this paper, we do not adopt a single definition of public procurement, as methodologically we chose to leave such distinctions to our interviewees who were encouraged to discuss whatever processes and components they personally and professionally associated with public procurement. 
Given the broader diversity of the term, as would be expected, we observed differences across municipalities in what types of acquisitions and activities participants deemed to fall under the umbrella of "procurement". For example, procurement departments often did not oversee governments' acquisition and use of free technologies, which we discuss further in Section \ref{sec:practices-people}.

\section{Cities' AI Use Cases}\label{apdx:ai-use-cases}

\begin{table*}[h!]
\small
\begin{tabular}{ |c|c| } 
    \hline
  \textbf{Type of AI technology} & \textbf{Examples} \\ \hline
  Facilitating resident communication & Translation services, chatbots, 311 assistance, public meeting summaries \\ \hline
  Law enforcement & License plate readers, gunshot detection, object detection \\ \hline
  Smart cities/urban planning & Sensors to track service utilization, accident tracking, snow plow routing \\ \hline
  Assisting bureaucratic decision-making & Funding allocation, service allocation, school bus routing \\ \hline 
  Workplace productivity tools & Chatbots, image generation, voice generation, coding assistants \\

 \hline
\end{tabular}
\caption{We grouped the AI systems that municipal employees discussed procuring or adopting in interviews, into 5 categories based on their intended usage. We provide anonymized examples of types of AI systems that were mentioned in each category. Employees in each city shared at least one example that they were aware of belonging to one of these five categories.}
\label{table:ai_use_cases}
\end{table*}

Table \ref{table:ai_use_cases} groups examples of AI adopted by municipalities into five categories based on their intended usage. In our discussions, employees in each city shared at least one example that they were aware of belonging to one of these five categories.

Interestingly, not all of the employees that we interviewed were aware that other employees in their city had already procured or adopted AI technologies: for example, one city employee stated that to their knowledge, their city ``has never purchased anything AI related", whereas their colleagues stated that the city in fact has. 

\newpage
\section{Timeline of Events}
Our research on AI procurement, like previous empirical research, captures a particular and unique ``snapshot in time'' ~\cite{autio2023snapshot}. Below, we provide an abbreviated timeline of key events that occurred while we conducted our research:

\begin{itemize}
    \item \textbf{December 2023}: Interviews with local government employees begin.
    \item \textbf{March 2024}: The San José Government AI Coalition is publicly launched with a membership of over 150 agencies, releasing their first public policy templates for local governments. All U.S. local governments are invited to join. 
    
    Many (but not all) of our participating governments do decide to join, and consider how they might adopt the coalition's resources.
    \item \textbf{March 2024}: The Biden Administration issues a second RFI for M-24-18: ``Advancing the Responsible Acquisition of Artificial Intelligence in Government''. Informally referred to as the ``AI M-Memo'', the memo establishes requirements that federal agencies must follow when procuring AI systems, requiring additional impact assessments and governance requirements for ``rights- or safety-impacting AI''. The memo also outlines a clear vision to increase agency capacity for responsible AI innovation.

    Several participating employees take inspiration from the M-Memo when developing their own AI governance strategies. Our research group \href{https://www.cyber.pitt.edu/sites/default/files/AI/Procuring%20Public-Sector%20AI.pdf}{responds to the RFI (link)}, sharing learnings we've had since beginning the project.
    \item \textbf{June 2024}: Interviews with local government employees end.
    \item \textbf{October 2024}: The Biden administration adopts the AI M-Memo, issuing a series of short-term deadlines for federal agencies to comply. 
    \item \textbf{January 2025}: The Trump administration revokes the Biden Executive Order 14110 and the AI M-Memo on day one of holding office. Despite the orders being revoked, scholars reflect on the widespread impact their ideas have had for governments worldwide.
\end{itemize}

When writing our paper, we aimed to balance characterizing the rapidly-evolving AI-specific practices that governments have adopted, while trying to identify which findings felt most evergreen.
We theorize that any of the legacy, decades-old purchasing practices discussed in Section \ref{sec:findings-one} will continue to shape how AI procurement occurs for years to come. 
However, because interviewed cities were in early stages (their first year) of standing up new AI governance practices like AI-specific risk assessments (Section \ref{sec:new-practices}), we acknowledge many of these practices and their implications for AI vendors' compliance are likely to change as cities continue to iterate and share best practices. 
Our hope is to surface timely opportunities for the FAccT community to contribute to shaping the future of this rapidly evolving landscape.

\newpage
\section{Cities' emerging AI governance practices (extended)}\label{apdx:risk-assessments}

In this section, we present an extended description of interviewed governments' emerging AI governance practices. 
We do not provide exact descriptions to preserve the anonymity of participating cities, and instead aim to pull out broad trends across cities.

\paragraph{AI risk or impact assessments} Participants conducted additional risk or impact assessments to better understand the possible positive or negative impacts of procured AI systems. While some cities conducted such assessments in an informal or ad-hoc way, others had started to standardize assessment processes by creating assessment templates with lists of questions and considerations. Different cities also conducted risk or impact assessments at different phases of the procurement process: some assessment instruments could be completed based on a "purpose statement" for AI, before a specific vendor or AI system is identified. In contrast, other risk assessments can only be completed once a concrete system has been identified, \eg{} they require knowledge of the system's performance.

The role and purpose of these assessments varied across interviewed cities. In many cities, the risk assessment had no immediate outcome, but employees were encouraged to take action to manage and if possible, mitigate potential risks identified in the process. Beyond informing mitigation steps, some participants also used risk assessments to triage AI solutions into "high" or "low" risk categories, which then determined subsequent requirements for review and oversight. For example, one city required high-risk AI to have additional reporting requirements, further risk assessment, usage protocols, and regular post-deployment monitoring. Participants viewed risk triaging as a way to reduce reviewing burden and better allocate their limited technical expertise. One participant who conducted AI risk assessments explained:

\begin{quote}
"[When triaging risk], we're just trying to get a sense of how thorough a review we need to do, because we're working with very limited capacity and resources. So we've got to decide: is this a low-risk system that we can just do a really quick look at? Or is this going to be something really sensitive and safety-impacting, rights-impacting, that we need to dedicate a lot of our time to?"   
\end{quote}

Participants also noted how risk triaging was also time-saving for their colleagues on the other end trying to purchase the AI, as put by one employee: "If it's low risk, I'll approve it, and you'll be on your way tomorrow!"

\paragraph{Practice vs. policy?} In many cities, employees made  changes to their procurement practices simply by adjusting their existing practices, \eg{} by electing to include vendor reporting requirements in an AI RFP.  Some cities decided to make these changes in their practices more formal or mandatory for vendors or city employees, by adopting policies or passing laws that required them. For example, Rebecca, an IT department leader walked through how their city's formal AI policy spelled out mandatory steps, such as a risk assessment, that city employees must complete for any AI procurement. The participant viewed the policy, which was passed by their city council, as an "accountability trigger" to incentivise compliance for both colleagues and vendors:

\begin{quote}
    "Council adopted the policy. So you can't just say no. I'm going to have some leverage to say, we can't just say we're not going to do this. [...] [The policy] is really meant to be a way to say the city is going to be taking this on, these are our values." 
\end{quote}

Participants in another city shared that while ideally someday they would like to institutionalize their practices via a formal policy, at the time of interviewing, they did not yet have one:

\begin{quote}
    "We very intentionally have not put out a [formal] AI policy yet, because we wanted more [community and government] input on it. And the space, especially in 2023, was very new for us. So we wanted to get a better understanding before asking our leadership to pass a policy."
\end{quote}

This city has since adopted a formal AI policy following engagement with the community, experts, and agency staff.

\newpage

\newpage
\section{Interview Protocol}\label{apdx:protocol}

We began the interview by reminding the participant of our informed consent protocol (approved by our institution's IRB board), and asking for their consent to record.

\subsection{Introduction} 
The goal of this interview is to learn more about existing procurement practices specifically for artificial intelligence, or AI, technologies in your city. We adopt a wide definition of AI as "any machine-based system that can make predictions, recommendations, or decisions". This would include technologies such as facial recognition, gunshot recognition technology, resume screening technology, ChatGPT, etc. 

Our goal is not to assess your practices, but rather to identify needs and opportunities for researchers as partners to support US cities.

\textbf{Q1.1:} Can you tell me a bit about your current role, and any past work experiences or responsibilities relating to artificial intelligence?

\textbf{Q1.2:} Have you ever been involved in a past procurement of an artificial intelligence technology?
\begin{itemize}
    \item If YES: How were you involved?
    \item If NO: Has your [agency] ever considered or talked about procuring AI?
\end{itemize}

\subsection{Walk-through} 
\emph{The goal is to understand how a "typical" AI procurement occurs in the city. Our goal is not to impose structure on the participant's description, but rather allow them to describe how they personally view/understand the procurement process.}

If it doesn't come up naturally, we can prompt the participant to reflect on specific parts of procurement, such as (1) Planning, (2) RFP writing, (3) Evaluating Vendors, (4) Contracting, (5) Designing/Building/Evaluating the AI, and (6) Deployment, and (7) Post-deployment.

\textbf{Q2.1., Walkthrough.} Can you briefly walk us through how a typical procurement involving an artificial intelligence technology would occur in your city? We're specifically interested in understanding any difference between a standard technology procurement, vs. a procurement involving AI.

\begin{itemize}
    \item If never procured AI: \eg{} imagine your city is considering procuring an enterprise-level generative AI product, like a chatbot to screen 311 questions.
\end{itemize}

Drill-down prompts on specific parts of the procurement process:

Planning (Problem Formulation): 
\begin{enumerate}
    \item What does your city do to plan for the procurement before the RFP (request for proposal) writing stage?
    \item (if not covered) Pre-RFP, how does the agency identify that an AI tool might be a part of the solution (rather than a tool that does not use AI)?
    \item (if not covered) Do you have a process for evaluating the risks of a proposed AI technology before RFP writing?
    \begin{itemize}
        \item If YES: What about potential mitigation processes for these risks?
    \end{itemize}
\end{enumerate}

RFPs:
\begin{enumerate}
    \item Is there anything different in the content of the RFP for AI procurements, compared to standard technology procurements that do not involve AI?
    \item (if not covered) In the RFP, do you ask vendors questions about potential risks and mitigation strategies?
\end{enumerate}

Evaluating Proposals:
\begin{enumerate}
    \item How does your city evaluate proposed AI solutions? 
We are especially interested in differences between evaluating standard technology vs. AI proposals.

    \item (if not covered) What information do you ask vendors to report in their proposal? Do you ever encounter "trade secrecy" claims?

    \item (if not covered) What measures do you expect them to report?
Do they validate that the technology works as claimed using data from your city?
\end{enumerate}

Contracting:
\begin{enumerate}
    \item Are there any differences in the contracting process for AI vs. non-AI (standard technology) technologies?
    \item Are there specific terms and conditions that you include in AI contracts?
    \item Can you share a past contract for an AI technology with us?

\end{enumerate}

AI Design, Development, and Evaluation: 
\begin{enumerate}
    \item How are people from your city involved with the design, development and evaluation of AI technologies under contract?
    \begin{itemize}
        \item If YES: How were you involved? What type of feedback did you give?
    \end{itemize}
    \item How often do vendors make changes to their technologies (like updating or improving it using data from your city) before they are deployed?
    \item How do vendors evaluate the AI solution they have designed and developed to make sure it fits your use case?
    \begin{itemize}
        \item Do they use data from your municipality for evaluation?
        \item What kind of measures do they look at and report to you?

    \end{itemize}
\end{enumerate}

AI Deployment:
\begin{enumerate}
    \item How often do vendors (or the city) provide training or onboarding for people who will be using the AI?
    \item How are agency workers involved in deciding the way the AI is used in their everyday practice?

\end{enumerate} 

Post-deployment: 
\begin{enumerate}
    \item How do you oversee and monitor deployed AI technologies?
    \begin{itemize}
        \item What is the vendors' responsibility?
        \item What if something goes wrong? (liability)
    \end{itemize}
\end{enumerate}

\textbf{Q2.3 (if unclear)} Can you remind me of who in your city is involved or oversees each phase of this procurement process?

\textbf{Q2.4 (if unclear)} Do you believe the process that we just went through together is representative of most AI procurements in your city (if relevant: beyond that specific example)?

\textbf{Q2.5.} Are there any existing policies in place that target the procurement of AI technologies specifically?
\begin{itemize}
    \item If YES: 
    \begin{itemize}
        \item Can you share your city's policies/guidelines with us?

        \item How long have these policies been in place?

    \end{itemize}
    \item If NO: 
    \begin{itemize}
        \item Is this something you anticipate being developed in the near future, or something that has been discussed?

    \end{itemize}
\end{itemize}

\textbf{Q2.6.} Can you direct us to your city's general procurement policies that may be applicable to AI technologies? e.g., such as data privacy policies?

\textbf{Q2.7.} Are there any AI technologies that come to be used through processes outside of the traditional procurement process? (e.g., research partnerships, foundations, donations, or free tools?)
\begin{itemize}
    \item Do these technologies undergo a similar "vetting" process to procured technologies?
    \begin{itemize}
        \item Do similar people evaluate these proposals?
        \item Do similar people oversee or monitor their deployment?
    \end{itemize}
\end{itemize}

\textbf{Q2.8.} Does your city consider opportunities to engage with residents who may be affected by an AI tool during the procurement process?

\subsection{Challenges \& Desires} 
\emph{The goal is to understand the participant's needs and desires to improve the procurement process.}

For the last part of our interview, we'd like to understand your opinions and wishes for improving AI procurement.

\textbf{Q3.1.} What do you believe are the main challenges or "pain points" for AI procurement in your city?
\begin{itemize}
    \item Do you have any suggestions as to how cities could improve their procurement of AI?

    \item (if relevant) Do you have any examples where [this challenge] happened in the past?

\end{itemize}

\textbf{Q3.2.} Can you imagine any new resources that could help you address these challenges?
\begin{itemize}
    \item What resource format would be most helpful?
ex: Checklists? Templates? Trainings?
\end{itemize}

\end{document}